\documentclass[12pt]{article}
\usepackage{epsf,amsfonts,latexsym,psfrag}
\newsymbol\ltimes 226E
\newsymbol\rtimes 226F
\epsfverbosetrue
\def\double{\mathbb}
\def\ccal{\cal}

\def\cc{{\double C}}

\def\rr{{\double R}}
\def\zz{{\double Z}}
\def\qqq{{\double Q}}
\def\llll{{\double L}}

\def\aa{{\cal A}}
\def\ccc{{\cal C}}
\def\dd{{\cal D}}

\def\hh{{\cal H}}
\def\hhh{{{\double H}}}

\def\mm{{{\ccal M}}}

\def\aa{{\cal A}}
\def\dd{{\cal D}}
\def\hh{{\cal H}}

\def\lll{{\cal L}}
\def\sss{{\cal S}}
\def\jj{{\cal J}}
\def\t{{\rm tr}\,}

\def\ddd{{\,\hbox{$\partial\!\!\!/$}}}

\def\de{\hbox{\rm d}}

\def\pa{{\partial}}

\def\ot{\otimes}
\def\op{\oplus}

\def\bb{\begin{eqnarray}}
\def\ee{\end{eqnarray}}
\def\eee{\nonumber\end{eqnarray}}
\def\pp{\pmatrix}
\def\qq{\quad}

\begin{document}

\hsize 17truecm
\vsize 24truecm
\font\fifteen=cmbx10 at 15pt
\font\twelve=cmbx10 at 12pt
\font\eightrm=cmr8
\baselineskip 18pt

\begin{titlepage}
\setcounter{footnote}{0}
\renewcommand{\thefootnote}{\arabic{footnote}}

\centerline{\twelve CENTRE DE PHYSIQUE TH\'EORIQUE
\footnote{
Unit\'e Propre de Recherche 7061}}
\centerline{\twelve CNRS - Luminy, Case 907}
\centerline{\twelve 13288 Marseille Cedex 9}
\vskip 3truecm

\centerline{\fifteen A FAREWELL TO UNIMODULARITY}

\vskip 2cm

\begin{center}
{\bf Serge LAZZARINI}
\footnote{\, also at Universit\'e de la
M\'editerrann\'ee (Aix-Marseille II),
\texttt{sel@cpt.univ-mrs.fr}}
\hskip 0.2mm and   {\bf Thomas SCH\"UCKER}
\footnote{\, also at Universit\'e de Provence
(Aix-Marseille I), 
\texttt{schucker@cpt.univ-mrs.fr}}
\end{center}

\vskip 1.5cm

\hfill{\em To the memory of Daniel TESTARD}
\vskip 2cm
\centerline{\bf Abstract}

\medskip

We interpret the unimodularity condition in almost
commutative geometries as central extensions of spin
lifts. In Connes' formulation of the standard model this
interpretation allows to compute the hypercharges of
the fermions.

\vskip 2truecm
 PACS-92: 11.15 Gauge field theories\\
\indent MSC-91: 81T13 Yang-Mills and other gauge
theories

\vskip 1truecm

\noindent march 2001
\vskip 1truecm
\noindent CPT-01/P.4193\\
\noindent hep-th/yymmxxx

\vskip1truecm

 \end{titlepage}

\section{Introduction}

Connes' noncommutative geometry
\cite{book}\cite{grav} allows to derive the intricate
rules of the  Yang-Mills-Higgs model building kit from
first principles
\cite{tresch}\cite{grav}\cite{cc}: in almost
commutative geometries, certain Yang-Mills-Higgs
forces appear as pseudo forces associated to gravity,
just as in Minkowskian geometry, the magnetic force
appears as a pseudo force associated to electricity. The
input of the Yang-Mills-Higgs kit, a compact group,
three unitary representations and a certain number of
coupling constants, is drastically constrained by almost
commutative geometry and the simplest possible input
is the standard model for leptons. To be more concrete:
the following properties of the complete standard
model are {\it ad hoc} in the Yang-Mills-Higgs kit, but they
derive from noncommutative geometry:
\begin{itemize}
\item Fermions transform according to fundamental or
trivial representations under isospin and colour,
\item parity violation is explicit, not spontaneous
\cite{florian},
\item strong forces couple vectorially,
\item colour is unbroken,
\item isospin is broken spontaneously by one
 doublet of scalars with hypercharge -1/2,
\item the gauge group must contain a non-vectorial $U(1)$ 
\cite{gis}.
\end{itemize}
In the past, quite some effort has been
devoted to an understanding of charge quantization,
e.g. magnetic monopoles, anomalies, grand unification. In the
standard model, charge quantization follows from an
even more intriguing property, O'Raifeartaigh's
$\zz_2\times\zz_3$ reduction \cite{or}. Noncommutative
geometry only allows algebra representations for
fermions and therefore restricts their
$U(1)$ charges to
$-1$, 0, 1. However the unimodularity condition,
presently the only kill-joy of the geometric formulation of
the standard model, always stood in the way to charge
quantization. Here we reconcile the unimodularity
condition with Connes' derivation of the Yang-Mills-Higgs
kit. For the standard model, this reconciliation implies a
strong form of charge quantization.

\section{Commutative geometry, gravity
and electromagnetism}

Let us summarize the punch line of Connes' geometry.
Let
$M$, `spacetime', be a compact, Riemannian spin
manifold and
$\aa=\ccc^\infty(M)$ the commutative, associative
algebra of differentiable functions from $M$ into
$\cc$. In a first step, Connes reformulates the
geometry of spacetime in algebraic terms without
using the commutativity of the algebra. Basic
ingredients of his formulation are: the algebra $\aa$,
its faithful representation by pointwise multiplication
on the Hilbert space $\hh=\lll^2(\sss)$ of square
integrable spinors, the self-adjoint Dirac operator
$\ddd$ on $\hh$, the charge conjugation or real
structure
$J=C\circ\,$complex conjugation, and in the four
dimensional case the chirality operator
$\chi=\gamma_5 $.

In a second step, he replaces the algebra
$\ccc^\infty(M)$ by a noncommutative algebra. If $M$
instead of describing spacetime was describing phase
space then the algebra of observables from quantum
mechanics would be an example of this
noncommutative algebra represented on the Hilbert
space of wave functions.

Connes' formulation is precise enough to allow
repeating Einstein's derivation of general relativity
using only the algebraic data $(\aa,\hh,\ddd,J,\chi )$
of a four dimensional spacetime. As Einstein's
derivation, Connes' consists of two strokes, kinematics
and dynamics for the gravitational field. Connes' first
stroke is the fluctuating metric
\cite{tresch}\cite{grav} where he uses essentially the
spin lift of algebra automorphisms to the Hilbert
space to identify the Dirac operator as gravitational
field. The second stroke is the celebrated spectral action
\cite{cc}, that reproduces the Einstein-Hilbert action
from the spectrum of the Dirac operator $\ddd$.

In the commutative case, $\aa=\ccc^\infty(M)$, the
group of automorphisms is just the group of
diffeomorphisms, Aut$(\aa)$=Diff$(M)$. We interpret
a diffeomorphism
$\varphi $ {\bf locally} as general coordinate
transformation. The receptacle group of
automorphisms of the algebra lifted to the Hilbert space is
\bb{\rm Aut}_\hh(\aa):=\{U\in {\rm End}(\hh),\
UU^* =U^*U=1,\ UJ=J U,\ U\chi =\chi U,\ 
i_U\in{\rm Aut}(\rho (\aa))\},\ee
with $i_U(x):=UxU^{-1}$. The first three properties
say that a lifted automorphism $U$ preserves probability,
charge conjugation and chirality. The
fourth, called {\it covariance property}, allows to
define the projection
$p:\ {\rm Aut}_\hh(\aa)\longrightarrow {\rm
Aut}(\aa)$ by
\bb p(U)=\rho ^{-1}i_U\ee
In our
case, a local calculation yields
${\rm Aut}_{\lll^2(\sss)}(\ccc^\infty (M))={\rm
Diff}(M)\ltimes \,^MSpin(4)$. We say receptacle because
already in six dimensions,
${\rm Aut}_{\lll^2(\sss)}(\ccc^\infty (M))$ is larger
than ${\rm Diff}(M)\ltimes \,^MSpin(6)$.
 We still have to construct the lift
$L: {\rm Diff}(M)\longrightarrow\,^MSpin(n)$ with
$p(L(\varphi ))=\varphi $.  In the coordinates $\tilde
x^{\tilde \mu}
$, the spin lift $L(\varphi )$ applied to a spinor
$\psi (\tilde  x)\in\lll^2(\sss)$ takes the explicit form
\cite{lift},
\bb \left( L(\varphi )\psi \right) (
x)=\left.S\left(\Lambda (\varphi
)\right)\right|_{\varphi ^{-1}( x)}\psi
({\varphi^{-1}( x)}),\label{spi}\ee 
with $x=\varphi
(	\tilde x)$, the local Lorentz transformation
\bb\left.\Lambda(\varphi)\right|_{\tilde x}
=\left.\sqrt{\jj^{-1T}\tilde g\jj^{-1}}\right|_{\varphi
(\tilde x)}\left.\jj
\right|_{\tilde x}\left.\sqrt{\tilde g^{-1}}\right|_{\tilde
x},\ee the matrix
$\tilde g_{\tilde \mu \tilde \nu }(\tilde x)=\tilde
g(\pa/\pa \tilde x^{\tilde \mu} ,\pa/\pa \tilde x^{\tilde
\nu})$ of the metric in the coordinates
$\tilde x$, the Jacobian of the diffeomorphism
${\jj(\tilde x)^{-1\tilde \mu }}_{ \mu} =\pa \tilde
x^{\tilde \mu }/\pa \ x^{ \mu }$ and the spin lift
\bb S: SO(4)&\longrightarrow& Spin(4)\cr
\Lambda =\exp\omega &\longmapsto&\exp \left[
{\textstyle\frac{1}{4}} \omega _{ab}
\gamma ^{a}\gamma ^b\right] \label{spin},\ee 
$\omega
=-\omega ^T\,\in so(4)$.
Now we can characterize algebraically the spin group in any
dimension $n$ of spacetime $M$ as image of the
automorphism group under the lift: $^MSpin(n)=L({\rm
Aut}(\ccc^\infty (M)))\subset {\rm
Aut}_{\lll^2(\sss)}(\ccc^\infty (M))$.
 The spin lift $L$ is of course
double-valued. In $n=4$ dimensions, this double-valuedness is
accessible to quantum mechanical experiments, e.g. neutrons
have to be rotated through an angle of $720^{\circ}$  before
interference patterns repeat \cite{neu}.

Consider the flat Dirac operator
$\tilde\ddd$  in inertial coordinates $\tilde x^{\tilde
\mu}$,
\bb \tilde \ddd =i{\delta ^{\tilde \mu} }_a\gamma
^a{\,\frac{\pa}{\pa \tilde x^{\tilde \mu} }} .\ee 
We have
written
${\delta ^{\tilde \mu }}_a\gamma ^a$ instead of
$\gamma^{\tilde \mu}$ to stress that the selfadjoint
$\gamma$ matrices are $\tilde x$-independent. Let us
`fluctuate' it by changing to general (curved) coordinates $x =
\varphi (\tilde x)$:
\bb L(\varphi )\tilde \ddd L(\varphi )^{-1}
=:\ddd= i\hbox{$
e^{-1\,\mu}$}_a\gamma^a\left[
\,\frac{\pa}{\pa x^{\mu}}\,+s(\omega
_{\mu})\right],\label{2}\ee 
where
$ e^{-1}=\sqrt{\jj \jj^T}$ is a symmetric matrix,
\bb s: so(4)&\longrightarrow& spin(4)\cr
\omega &\longmapsto&  {\textstyle\frac{1}{4}} \omega
_{ab}
\gamma ^{a}\gamma ^b \ee 
is the Lie algebra isomorphism
corresponding to the lift (\ref{spin}) and
\bb 
\omega_{\mu}(
x)=\left.\Lambda\right|_{\varphi^{-1}( x)}
\,\frac{\pa}{\pa x^{ \mu}}\,
\left.\Lambda^{-1}\right|_{ x}\,.\ee 
For the `spin
connection' $ \omega$ we recover
 the well known expression
\bb{\omega^a}_{b \mu}( e)=
{\textstyle\frac{1}{2}}\left[( \pa_{
\beta}\hbox{$ e^a$}_{ \mu})- (
\pa_{\mu}\hbox{$ e^a$}_{ \beta})+
\hbox{$ e^m$}_{ \mu}(
\pa_{\beta  }
\hbox{$ e^m$}_{
\alpha})\hbox{$ e^{-1\,\alpha}$}_a
\right]\hbox{$ e^{-1\,
\beta}$}_b\ -\ [a \leftrightarrow b]\ee
 of the torsionless spin connection in terms of the first
derivatives of
$ e.$ Modulo flatness (encoded in the constraint
that
$ e^{-1}=\sqrt{\jj \jj^T}$ is not a general positive
matrix) the fluctuation of the flat Dirac operator
produces the general one and the kinematics of the
gravitational field is the set of positive matrices $
e$ with smooth spacetime dependence or equivalently
the set of all torsionless Dirac operators.

The second stroke is the spectral action, a
functional on this latter set, that defines the dynamics
of the gravitational field. The beauty of Chamseddine
\& Connes' approach \cite{cc} to general relativity is
that it works precisely because the Dirac operator
$\ddd$ plays two roles simultaneously, it defines the
dynamics of matter and it parameterizes the set of all
Riemannian metrics.
Their starting point is the
simple remark that the spectrum of the Dirac operator
is invariant under diffeomorphisms interpreted as
general coordinate transformations. From
$\ddd\chi=-\chi\ddd$ we know that the
spectrum of
$ \ddd$ is even. We may therefore consider only
the spectrum of the positive operator
$\ddd^2/\Lambda^2$ where we have divided by a fixed
arbitrary energy scale to make the spectrum
dimensionless. If it was not divergent the trace $\t
\ddd^2/\Lambda^2$ would be a general
relativistic action functional. To make it convergent,
take a differentiable function
$f:\rr_+\rightarrow\rr_+$ of sufficiently fast
decrease such that the action
\bb S_{CC}:=\t f( \ddd^2/\Lambda^2)\ee
converges. It is still a diffeomorphism invariant
action. Using the heat kernel expansion it can be
computed asymptotically:
\bb S_{CC}=
\int_M
[\Lambda_c-{\textstyle\frac{m_P^2}{16\pi}}R
+a(5\,R^2-8\,{\rm Ricci}^2-7\,{\rm
Riemann}^2)]\,\sqrt{\det g_{\mu\nu}}\de^4x \,+\,
O(\Lambda^{-2}),\ee where the cosmological constant
is $\Lambda_c=
{\textstyle\frac{f_0}{4\pi^2}}\Lambda^4$, the Planck
mass is
$m_P^2={\textstyle\frac{f_2}{3\pi}}\Lambda^2$ and
$a={\textstyle\frac{f_4}{5760\pi^2}}$. The
Chamseddine-Connes action is universal in the sense
that the `cut off' function $f$ only enters through its
first three `moments', $f_0:=\int_0^\infty uf(u)\de u$,
$f_2:=\int_0^\infty f(u)\de u$ and $f_4=f(0)$. Thanks
to the curvature  square terms the
Chamseddine-Connes action is positive and has
minima. For instance the 4-sphere with a radius of
$(11f_4)^{1/2}(90\pi
(1-(1-11/15\,f_0f_4f_2^{-2})^{1/2}))^{-1/2}$
 times the Planck length is a ground state. This
minimum breaks the diffeomorphism group
spontaneously down to the isometry group $SO(5)$.
The little group consists of those lifted
automorphisms that commute with the Dirac operator
$ \ddd$. Let us anticipate that the spontaneous
symmetry breaking of the Higgs mechanism will be a
little brother of this gravitational break down.
Physically the gravitational symmetry breaking seems
to regularize the initial cosmological singularity.

As a bonus, this algebraic derivation of gravity
achieves the unification with electromagnetism by a
straight forward central extension of the lift
$L$. Since
particles and antiparticles have opposite charge, we
have to separate them before turning on the
electromagnetic field. Technically this is done by
doubling the Hilbert space,
\bb \aa=\ccc^\infty(M)\owns a,&
\hh_t=\lll^2(\sss)\ot\cc^2\owns\psi _t=\pp{\psi \cr
\psi ^c},&
\rho _t(a)=\pp{a1_4&0\cr 0&\bar a1_4},\label{tens1}\\
\dd_t=\pp{\ddd&0\cr 0&\ddd},&J_t=C\ot\pp{0&1\cr
1&0}\circ\,{\rm c\ c},&\chi _t=\gamma
_5\ot\pp{1&0\cr 0&1}.\label{tens2}\ee
 We emphasize that $\psi ^c$
is not a new degree of freedom but we will make $\psi
^c$ the antiparticle of
$\psi $ {\it at the end} by imposing $J_t\psi _t=\psi _t.$
This disentangling of particles and antiparticles is Dirac's
reinterpretation of the antiparticles as holes.  Now ${\rm
Aut}_{\hh_t}(\ccc^\infty (M))={\rm Diff}(M)\ltimes
\,^M(Spin(4)\times U(1))$ and we wish to extend the lift
$L$ to include the $U(1)$ gauge transformations. At
this point we anticipate that we aim at
noncommutativity: for a noncommutative
algebra $\aa$ there is a strong link between its group
of automorphisms Aut$(\aa)$ and its group of
unitaries
\bb U(\aa):=\left\{ u\in \aa,\ uu^*=u^*u=1\right\}.\ee
In our present commutative example $\ccc^\infty(M)$,
this link is not visible, but its group of unitaries
consists precisely of $U(1)$ gauge transformations,
$U(\ccc^\infty(M))=\,^MU(1)$.
Furthermore there is a natural class of centrally
extended lifts $\llll$ on $\aa=\ccc^\infty(M)$:
\bb\llll= (L,\ell):{\rm Aut}(\aa)\ltimes U(\aa)
&\longrightarrow&\,^M\left(Spin(4)\times
U(1)\right)\
\subset\ {\rm Aut}_{\hh_t} (\aa)\cr
(\varphi,u)&\longmapsto&
\llll(\varphi ,u)=\left(L(\varphi),\ell(u)\right)\eee
with
\bb
\ell(u)=\rho_t(u^{q/2})J_t\rho_t(u^{q/2})J_t^{-1},\qq
q\in 2\zz\ \ {\rm or}\ q\in\qqq.\label{ext}\ee
We may
allow rational charges $q$ if we do admit spin
representations, i.e. multi-valued representations. Let us again
fluctuate the flat Dirac operator $\tilde \dd_t$ and compute
the spectral action of the fluctuated Dirac operator:
\bb\llll(\varphi ,u)\tilde \dd_t\,
\llll(\varphi ,u)^{-1}=
\pp{ \ddd&0\cr 0&{C\ddd C^{-1}}}.\ee
As before, a straight forward calculation yields the
covariant derivative:
\bb \ddd=
i\hbox{$ e^{-1\,\mu}$}_a\gamma^a\left[
\pa_{ \mu}+s(\omega
_{\mu})-q A_{\mu}\right],
\qq A_{\mu}=u\pa/\pa x^{ \mu }\,u^{-1}, \ee
identifying indeed the power $q$ in the central
extension (\ref{ext}) as electric charge. The second
stroke unifies gravity with electrodynamics:
\bb S_{CC}&=&\t f(\tilde \ddd^2/\Lambda^2)\cr\cr  &=&
\int_M
[\Lambda_c-{\textstyle\frac{m_P^2}{16\pi}}R
+a(5\,R^2-8\,{\rm Ricci}^2-7\,{\rm
Riemann}^2)\cr
&&\qq\qq\qq\qq\qq\qq\qq\qq\qq\qq\qq
+{\textstyle\frac{1}{4g^2}}F_{\mu
\nu }^*
 F^{\mu \nu }
]\,\sqrt{\det g_{\mu\nu}}\de^4x \,+\,
O(\Lambda^{-2}),\ee
where the electric coupling  constant
is $g^2={\textstyle\frac{6\pi ^2}{f_4}} $.
Although the algebra of functions on spacetime is
commutative, its group of automorphisms, the
diffeomorphism group, and its spin lift, the local
Lorentz group,  are nonAbelian. Consequently general
relativity has nonlinear field equations. On the other
hand, the group of unitaries remains Abelian and
Maxwell's equations are linear. Therefore the electric
charge $q$ and the electric coupling constant
$g={\epsilon_0}^{-1/2}$ only appear as products $qg$
and by means of a finite renormalization of the coupling
constant, we may put
$q=1$ for the electron.

\section{Lifts in finite, noncommutative
geometries and their central extensions}

The algebra $\aa$ is a real, associative involution
algebra with unit, that admits a faithful 
* representation $\rho $. In
finite dimensions, a simple such algebra is a real, complex or
quaternion matrix algebra,
$\aa= M_n(\rr),\ M_n(\cc)$ or
$M_n(\hhh) $, represented irreducibly on the Hilbert
space
$\hh=
\rr^n,\ \cc^n$ or $\cc^{2n}$. In the first and third
case, the representations are the fundamental ones,
$\rho(a)=a,\ a\in\aa$, while $M_n(\cc)$ has two
non-equivalent irreducible representations on
$\cc^n$, the fundamental one, $\rho(a)=a$ and its
complex conjugate $\rho(a)=\bar a$. In the general
case we have sums of simple algebras and sums of
irreducible representations. To simplify notations, we
concentrate on complex matrix algebras $M_n(\cc)$ in this
section. Indeed the others, $M_n(\rr)$ and $M_n(\hhh)$, 
do not have central unitaries close to the identity. In the
following it will be important to separate the commutative and
noncommutative parts of the algebra:
\bb\aa=\cc^M\oplus
\bigoplus_{k=1}^N M_{n_k}(\cc)\ \owns
a=(b_1,...b_M,c_1,...,c_N),\qq n_k\geq 2.
\label{algebra}\ee
Its group of unitaries is
\bb U(\aa)=U(1)^M\times
\matrix{N\cr \times\cr {k=1}} U(n_k)\
\owns\ u=(v_1,...,v_M,w_1,...,w_N)\ee
and its group of central
unitaries 
\bb U^c(\aa):=U(\aa)\cap\,{\rm
center}(\aa)=U(1)^{N+M}\
\owns\ u_c=
( v_{c1},...,v_{cM},w_{c1}1_{n_1},...,w_{cN}1_{n_N}).\ee
The component of the automorphisms group ${\rm
Aut}(\aa)$, that is
connected to the identity, is the group of inner
automorphisms,
${\rm Aut}(\aa)^e={\rm In}(\aa)$. There are additional,
discrete automorphisms, the complex conjugation and, if
there are identical summands in $\aa$, their permutations.
These discrete automorphisms do not concern us here. An
inner automorphism is of the form $i_u(a)=u a u^{-1}$ for
some unitary $ u \in U(\aa)$. 
Multiplying $u$ with a central unitary
$u_c$ of course does not affect the inner automorphism
$i_{u_cu}=i_u$. Note that this ambiguity distinguishes
between `harmless' central unitaries $v_{c1},...,v_{cM}$ and
the others, $w_{c1},...,w_{cN}$, in the sense that
\bb {\rm In}(\aa)=U^n(\aa)/U^{nc}(\aa),\label{inntrue}\ee
where we have defined the group of noncommutative unitaries
\bb U^n(\aa):=\matrix{N\cr \times\cr {k=1}} U(n_k)\
\owns\ w\ee
and $U^{nc}(\aa):=U^n(\aa)\cap U^c(\aa) \owns w_c$.
The map 
\bb i:U^n(\aa)&\longrightarrow&{\rm In}(\aa)\cr 
(1,w)&\longmapsto&i_w\ee
 has kernel Ker$\,i=U^{nc}(\aa)$. 

The lift of an inner
automorphism to the Hilbert space has a simple closed
form
\cite{tresch}, $L=\hat L\circ i^{-1}$ with
\bb \hat L(w)=\rho(1,w)J\rho(1,w)J^{-1}.\ee
It satisfies $p(\hat L(w))=i(w)$.
If the kernel of $i$ is contained in the kernel of $\hat L$ then
the lift is well defined, as e.g. for $\aa=\hhh$,
$U^{nc}(\hhh)=\zz_2$. 
\begin{eqnarray}
&&{\rm Aut}_\hh(\aa)\nonumber \\
&& \hskip -2mm p\ 
\parbox{6mm}{\begin{picture}(20,10)
\put(0,15){\vector(0,-1){30}}
\put(15,-15){\vector(-1,4){8}}
\put(15,-15){\vector(0,1){33}}
\end{picture}}
 L 
\parbox{8mm}{\begin{picture}(20,10)
\put(30,-15){\vector(-1,2){16}}
\put(32,-10){$\hat{L}$}
\end{picture}}
\parbox{12mm}{\begin{picture}(20,10)
\put(65,-15){\vector(-2,1){67}}
\end{picture}}
\ell
\\[4mm]
&&{\rm In}(\aa)\stackrel{i}{\longleftarrow} 
U^n(\aa)\begin{array}{c}\\[-3mm]
\hookleftarrow \\[-5mm]
\stackrel{\vector(1,0){15}}{\mbox{\footnotesize $\det$}} 
\end{array}
U^{nc}(\aa) \nonumber
\end{eqnarray}
For more complicated real or
quaternionic algebras, $U^{nc}(\aa)$ is finite and the lift $L$
is multi-valued with a finite number of values. For
noncommutative, complex algebras, their continuous family of
central unitaries can not be eliminated except for very special
representations and we face a continuous infinity of values.
The solution of this problem follows an old strategy: {\it `If
you can't beat them, adjoin them'.} Who is {\it them?} The
harmful central unitaries $w_c\in U^{nc}(\aa)$ and adjoining
means central extending.
 The central extension (\ref{ext}), generalizes naturally from
the algebra $\cc$ to our present setting:
\bb \ell(w_c)&=&\rho\!\! \left(
\prod_{j_1=1}^N(w_{cj_1})^{q_{1j_1}},
...,\prod_{j_M=1}^N(w_{cj_M})^{q_{Mj_M}},\right.&\cr &&
\left.\qq
\prod_{j_{M+1}=1}^N(w_{cj_{M+1}})^{q_{{M+1},j_{M+1}}}
1_{n_1},
...,\prod_{j_{M+N}=1}^N(w_{cj_{M+N}})^{q_{{M+N},j_N}}
1_{n_N}
\right)
 J
\rho (...)
\,J^{-1}
\label{ell}\ee
with the $(M+N)\times N$ matrix of charges $q_{kj}$. The
extension satisfies indeed
$p(\ell(w_c))=1\in{\rm In}(\aa)$ for all $w_c\in
U^{nc}(\aa)$. 

Having adjoined the harmful, continuous central unitaries, we
may now stream line our notations and write the group of
inner automorphisms as 
\bb {\rm In}(\aa)=\left( 
\matrix{{N} \cr \times \cr k=1} SU(n_k)\right) /\Gamma 
\owns[w_\varphi] = [(w_{\varphi 1},...,w_{\varphi N})]\ {\rm
mod}\
\gamma \label{innfake} .\ee
$\Gamma $ is the discrete group
\bb\Gamma=\matrix{{N}\cr \times\cr
{k=1}}\zz_{n_k}\ \owns\ (z_11_{n_1},...,z_N1_{n_N}),\qq
z_{k}=\exp[-m_{k}2\pi i/n_k],\ m_k=0,...,n_k-1
. \label{discrete}\ee
The quotient is factor by factor. This way to write inner
automorphisms is convenient for complex matrices, but not
available for real and quaternionic matrices. Equation
(\ref{inntrue}) remains the general characterization of
inner automorphisms.

The lift $L(w_\varphi )=(\hat L\circ i^{-1})(w_\varphi )$
is multi-valued with, depending
on the representation, up to $ |\Gamma |=\prod_{j=1}^N n_j$
values. More precisely the multi-valuedness of $L$ is indexed
by the elements of the kernel of the projection $p$ restricted
to the image $L({\rm In}(\aa))$. Depending on the choice of
the charge matrix
$q$, the central extension $\ell$ may reduce this
multi-valuedness. Extending harmless central unitaries is
useless for any reduction. With
the multi-valued group homomorphism
\bb (h_\varphi,h_c) : U^n(\aa)&\longrightarrow & {\rm
In}(\aa)\times U^{nc}(\aa)\cr  
(w_j) & \longmapsto &((w_{\varphi j} , w_{cj}))=((w_j(\det
w_j)^{-1/n_j},(\det w_j)^{1/n_j}))\label{isom},\ee
 we can write the two lifts
$L$ and
$\ell$  
together in closed form
$\llll:U^n(\aa)\rightarrow
{\rm Aut}_\hh(\aa)$:
\bb\llll(w)&=&L(h_\varphi (w))\,\ell(h_c(w))\cr \cr 
&=&
\rho\!\! \left(
\prod_{j_1=1}^N(\det w_{j_1})^{\tilde q_{1j_1}},
...,\prod_{j_M=1}^N(\det w_{j_M})^{\tilde
q_{Mj_M}},\right.\cr &&\left.\qq
w_1\prod_{j_{M+1}=1}^N(\det w_{j_{M+1}})^{\tilde
q_{{M+1},j_{M+1}}}, ...,w_N\prod_{j_{N+M}=1}^N(\det
w_{j_{N+M}})^{\tilde q_{{N+M},j_{N+M}}}\right)
\nonumber\\[2mm]
&&\times\,
J \rho (...) J^{-1}.\ee
We have set 
\bb\tilde q:=\left( q-\pp{0_{M\times N}\cr\cr  1_{N\times
N}}
\right) \pp{n_1&&\cr &\ddots&\cr &&n_N}^{-1}.\ee
Due to the
phase ambiguities in the roots of the determinants, the
extended lift
 $\llll$ is multi-valued in general. It is single-valued if the
matrix
$\tilde q$ has integer entries, e.g.
$q=\pp{0\cr 1_N}$, then $\tilde q=0$ and 
$\llll(w)=\hat L(w)$. On the other hand $q=0$ gives
$\llll(w)=\hat L(i^{-1}(h_\varphi (w)))$, not always well
defined as already noted. Unlike the extension $\ell$ of general relativity, equation
(\ref{ext}), and unlike the map $i$, the extended lift $\llll$ is
not necessarily even. We do impose this symmetry
$\llll (-u)=\llll (u)$ which translates into
conditions on the charges, conditions that depend on the
details of the representation $\rho $. 

 The lift
$\llll$ is the `representation up to a phase' that we have
studied earlier in the case of the standard model \cite{ls}.
 Let us note that $\llll$ is not the most general lift. We could
have added the harmless central unitaries,  and, if the
representation
$\rho $ is reducible, we could have chosen different charge
matrices in different irreducible components.

\section{The standard model}

The internal algebra $\aa$ is chosen as to reproduce 
$SU(2)\times U(1)\times SU(3)$ as subgroup of
$U(\aa)$,
\bb \aa=\hhh\op\cc\op
M_3(\cc)\,\owns\,(a,b,c).\ee 
The internal Hilbert
space is copied from the Particle Physics Booklet
\cite{data},
\bb \hh_L&=&
\left(\cc^2\ot\cc^N\ot\cc^3\right)\ \op\ 
\left(\cc^2\ot\cc^N\ot\cc\right), \\
\hh_R&=&\left(\cc\ot\cc^N\ot\cc^3\right)\ 
\op\ \left(\cc\ot\cc^N\ot\cc^3\right)\ 
\op\ \left(\cc\ot\cc^N\ot\cc\right).\ee
 In each summand, the first factor
denotes weak isospin doublets or singlets, the second
denotes
$N$ generations, $N=3$, and the third denotes colour
triplets or singlets.
Let us choose the following basis
of the internal Hilbert space, counting fermions and
antifermions independently as explained in section 2,
$\hh=\hh_L\op\hh_R\op\hh^c_L\op\hh^c_R
=\cc^{90}$: 
\bb
& \pp{u\cr d}_L,\ \pp{c\cr s}_L,\ \pp{t\cr b}_L,\ 
\pp{\nu_e\cr e}_L,\ \pp{\nu_\mu\cr\mu}_L,\ 
\pp{\nu_\tau\cr\tau}_L;&\cr \cr 
&\matrix{u_R,\cr d_R,}\qq \matrix{c_R,\cr s_R,}\qq
\matrix{t_R,\cr b_R,}\qq  e_R,\qq \mu_R,\qq 
\tau_R;&\cr  \cr 
& \pp{u\cr d}^c_L,\ \pp{c\cr s}_L^c,\ 
\pp{t\cr b}_L^c,\ 
\pp{\nu_e\cr e}_L^c,\ \pp{\nu_\mu\cr\mu}_L^c,\ 
\pp{\nu_\tau\cr\tau}_L^c;&\cr\cr  
&\matrix{u_R^c,\cr d_R^c,}\qq 
\matrix{c_R^c,\cr s_R^c,}\qq
\matrix{t_R^c,\cr b_R^c,}\qq  e_R^c,\qq \mu_R^c,\qq 
\tau_R^c.&\eee
This is the current eigenstate basis, the representation
$\rho$ acting on
$\hh$ by
\bb \rho(a,b,c):= 
\pp{\rho_{L}&0&0&0\cr 
0&\rho_{R}&0&0\cr 
0&0&{\bar\rho^c_{L}}&0\cr 
0&0&0&{\bar\rho^c_{R}}}\ee
with
\bb\rho_{L}(a):=\pp{
a\ot 1_N\ot 1_3&0\cr
0&a\ot 1_N&},\qq
\rho_{R}(b):= \pp{
b 1_N\ot 1_3&0&0\cr 0&\bar b 1_N\ot 1_3&0\cr 
0&0&\bar
b1_N}, \label{repr1}
\ee\bb 
  \rho^c_{L}(b,c):=\pp{
1_2\ot 1_N\ot c&0\cr
0&\bar b1_2\ot 1_N},\qq
\rho^c_{R}(b,c) := \pp{
1_N\ot c&0&0\cr 0&1_N\ot c&0\cr
0&0&\bar b1_N}.  \label{repr2} 
\ee
At this point we can explain why only isospin
doublets and singlets and colour triplets and singlets
are allowed in the fermionic representation:
all other irreducible group representations cannot be
extended to algebra representation. While the
tensor product of two group representations is again a
group representation, the tensor product of two
algebra representations is not an algebra
representation.
 The
apparent asymmetry between particles and
antiparticles -- the former are subject to weak, the
latter to strong interactions -- disappears after
application of the lift $\llll$ with
\bb J=\pp{0&1_{15N}\cr 1_{15N}&0}\circ 
\ {\rm complex\ conjugation}.\ee
 For the sake of
completeness, we record the chirality as matrix
\bb \chi=\pp{-1_{8N}&0&0&0\cr 0&1_{7N}&0&0\cr
 0&0&-1_{8N}&0\cr 0&0&0&1_{7N} }.\ee
The
internal Dirac operator
\bb \dd=\pp{0&\mm&0&0\cr 
\mm^*&0&0&0\cr 
0&0&0&\bar\mm\cr 
0&0&\bar\mm^*&0}\ee
contains the fermionic mass matrix of the standard
model,
\bb\mm=\pp{
\pp{1&0\cr 0&0}\ot M_u\ot 1_3\,+\,
\pp{0&0\cr 0&1}\ot M_d\ot 1_3
&0\cr
0&\pp{0\cr 1}\ot M_e},\ee
with
\bb M_u:=\pp{
m_u&0&0\cr
0&m_c&0\cr
0&0&m_t},&& M_d:= C_{KM}\pp{
m_d&0&0\cr
0&m_s&0\cr
0&0&m_b},\\[2mm] && M_e:=\pp{
m_e&0&0\cr
0&m_\mu&0\cr
0&0&m_\tau}.\ee
From the booklet we know that all indicated fermion
masses are different from each other and that the
Cabibbo-Kobayashi-Maskawa matrix  $C_{KM}$ is
non-degenerate in the sense that  no quark is
simultaneously mass and weak current eigenstate.

In the commutative setting,
$\aa=\ccc^\infty(M)$, $\hh=\lll^2(\sss)$, the algebraic
formulation of the fact that the Dirac operator is a first order
differential operator reads $[[\ddd,\rho(a)],J\rho(\tilde
a)J^{-1}]=0$ for
all $a,\tilde a\in\aa$. In Connes' noncommutative
geometry this property becomes an axiom, `the first order
axiom',
$[[\dd,\rho(a)],J\rho(\tilde a)J^{-1}]=0$. In the example of
the standard model, this axiom entails the existence of a
gauge group that commutes with the electro-weak
interactions and with the fermionic mass matrix and whose
fermion representation is vectorial
\cite{reb}. One of the important features of Connes' coding
of geometry via `spectral triples' $(\aa,\hh,\dd,J,\chi )$ is
that they can be tensorized. In the case of two Riemannian
manifolds, this tensor product describes the direct product.
This tensor product also generalizes the one, that Connes used
to unify electromagnetism with gravity, equations
(\ref{tens1},\ref{tens2}). The tensor product of a
Riemannian geometry $M$ and a zero dimensional one, i.e.
with finite dimensional algebra and Hilbert space like the
internal space of the standard model, has as Dirac operator
$\dd_t=\ddd\ot 1\,+\,\gamma _5\ot\dd$, the free, massive
Dirac operator. Its fluctuations with $\llll$ produce the
minimal couplings to gravity and to the non-Abelian gauge
bosons, and the Yukawa couplings to the Higgs boson which
in the example of the standard model comes out to be an
isospin doublet, colour singlet with hypercharge
$-{\textstyle\frac{1}{2}} $. 
The spectral action $S_{CC}$ then yields \cite{cc}, in
addition to the gravitational action, the entire bosonic
action of the standard model including the
Higgs sector with its spontaneous symmetry breaking.
The constraints for the coupling constants, $g_2^2=
g_3^2=3 \lambda$ occur
because the Yang-Mills actions and the $\lambda |\Phi|^4$
term stem from the same heat kernel coefficient $f_4 a_4$. 

Let us go back to the standard model as a Yang-Mills-Higgs
model and suppose that god has given the fermionic
representation content of isospin and colour. The
hypercharges can then be chosen arbitrarily, five rational
numbers, $y_1, ...,y_5$. $y_1$ is the hypercharge
of the left-handed  quarks, $y_2$ of the left-handed leptons,
$y_3$ of the right-handed up-quarks, $y_4$ the
hypercharge of the right-handed down quarks and $y_5$ of
the right-handed leptons. The Lorentz force prohibits
massless particles with non-vanishing electric charge.
Therefore the hypercharge $y_2$ of the purely left-handed
neutrinos must be different from zero and by a finite
renormalization we can set $y_2=-1/2$. If we want left- and
right-handed particles to have the same electric charge, then
we must impose the three conditions
\bb y_3={\textstyle\frac{1}{2}} +y_1=y_1-y_2,\qq
y_4=-{\textstyle\frac{1}{2}} +y_1=y_1+y_2,\qq
y_5=-{\textstyle\frac{1}{2}} +y_2=2y_2.
\label{3cond}\ee 
The hypercharges are then completely
fixed by putting
$y_1=1/6$ which amounts to choose the electric charge
of the quarks. Let us summarize
nature's choice of the fermionic hypercharges,
\bb y_1\ =&{\textstyle\frac{1}{6}},\qq &
6y_1=1\ {\rm mod}\ 2\qq{\rm and}\qq 1\ {\rm mod}\
3,\cr  
y_2\ =&-{\textstyle\frac{1}{2}},\qq &
6y_2=1\ {\rm mod}\ 2\qq{\rm and}\qq 0\ {\rm mod}\
3,\cr 
y_3\ =&{\textstyle\frac{2}{3}},\qq &
6y_3=0\ {\rm mod}\ 2\qq{\rm and}\qq 1\ {\rm mod}\
3,\cr 
y_4\ =&-{\textstyle\frac{1}{3}},\qq &
6y_4=0\ {\rm mod}\ 2\qq{\rm and}\qq 1\ {\rm mod}\
3,\cr 
y_5\ =&-1,\qq &
6y_5=0\ {\rm mod}\ 2\qq{\rm and}\qq 0\ {\rm mod}\
3.\label{nat}\ee
At this point O'Raifeartaigh \cite{or} remarks that after
renormalizing the hypercharges by a factor 6, the
isomorphism 
$ U(n) \rightarrow  [SU(n)\times U(1)]/\zz_{n}$ 
induced by the multi-valued homomorphism
$h$, equation (\ref{isom}), extents to the fermion
representations for isospin, $n=2$, and colour, $n=3$.
Indeed $\rho (u)=(\det u)^z\,u,\ u\in U(n),\ z\in\zz,$
defines a representation of $U(n)$ and under the
isomorphism  (\ref{isom}) it induces the fundamental
representation of
$SU(n)$ with $U(1)$ charge $1+zn$:
\bb (\det u)^z\,u=u_c^{zn}u_c^{}u_\varphi =
u_c^{1+zn}u_\varphi .\ee
The $U(n)$ representation $\rho (u)=(\det u)^z=u_c^{zn}$
induces the $SU(n)$ singlet representation with $U(1)$
charge $zn$. $U(1)$
charges are one modulo $n$ for fundamental multiplets,
zero modulo $n$ for singlets, precisely nature's choice
(\ref{nat}). In other words nature only represents a
quotient of
$SU(2)\times U(1)\times SU(3)$ on fermions (and bosons).
This faithfully represented quotient is
$[SU(2)\times U(1)\times SU(3)]/[\zz_2\times\zz_3].$
The $\zz_n$s are the centers of the $SU(n)$s but they do act
on the $U(1)$ which is not the case of $\Gamma $ in
(\ref{disstan}) below. O'Raifeartaigh's reduction is a
stronger restriction than charge quantization, the
hypercharges
$\times 6$ are not only to be integers, they must satisfy the
conditions in the second and third columns of equations
(\ref{nat}).

If we take the standard model as a noncommutative
geometry, 
isospin and colour of the fermions are given by the
geometry. Now what does this geometry tell us about the
hypercharges? The most general {\it algebra}
representation compatible with the first order axiom,
involves four `charges'
$\tilde y_1,...,\tilde y_4$. Each $\tilde y_\cdot$ can take
only 2 values,
$-1$ or $+1$,
\bb\rho_{L}(a):=\pp{
a\ot 1_N\ot 1_3&0\cr
0&a\ot 1_N&},\qq
\rho_{R}(b):= \pp{
b_31_N\ot 1_3&0&0\cr 0&b_4 1_N\ot 1_3&0\cr 
0&0&b_51_N},
\ee\bb 
  \rho^c_{L}(b,c):=\pp{
1_2\ot 1_N\ot c&0\cr
0&b_21_2\ot 1_N},\qq
\rho^c_{R}(b,c) := \pp{
1_N\ot c&0&0\cr 0&1_N\ot c&0\cr
0&0&b_21_N},  \ee
with $b_j:=[(1+\tilde y_j)b/2\,+\, (1-\tilde y_j)\bar b/2] $.
With the algebra automorphism of $\cc$,
$b\mapsto\bar b$, we can always arrange $\tilde
y_2=-1$. We must have $\tilde y_5=\tilde y_2$,
otherwise the right-handed leptons would be electrically
neutral leading to charged neutrinos in conflict with the
Lorentz force. On the other hand, we must have $\tilde
y_3\not=\tilde y_4,$ otherwise the bottom and top
masses would coincide after spontaneous symmetry
breaking. We are back at the representation of equations
(\ref{repr1},\ref{repr2}), possibly after a permutation of
$u$ with $d^c$ and of $d$ with $u^c$.

We have the following groups,
\bb U(\aa)=&\ SU(2)\times U(1)\times U(3)\qq& \owns
u=(u_0,v, w),\\
U^c(\aa)=&\ \zz_2\times U(1)\times U(1)\qq&\owns
u_c=(u_{c0},v_{c}, w_{c}1_3),\\
U^n(\aa)=&SU(2)\qq\times\qq U(3)\qq&\owns\qq\qq\ 
(u_0,w),\\
U^{nc}(\aa)=&\zz_2\qq\times\qq U(1)\qq&\owns\qq\qq\ 
(u_{c0},w_{c}1_3),\\
{\rm In}(\aa)=&\ [SU(2)\qq\times\qq SU(3)]/\Gamma
&\owns u_\varphi =(u_{\varphi 0}, w_{\varphi }),\\
\Gamma =&\zz_2\qq\times\qq \zz_3\qq&\owns\gamma = 
(\exp[-m_02\pi i/2],\exp[-m_22\pi i/3]),\label{disstan}
\ee
with
$m_0=0,1$ and $m_2=0,1,2$. 
Let us compute the receptacle
of the lifted automorphisms,
\bb {\rm Aut}_{\hh}(\aa)=
[U(2)_L\times U(3)_c&&\!\!\!\!\!\!\!\times U(N)_{qL}
\times U(N)_{\ell L}\times U(N)_{uR}
\times U(N)_{dR}]/[U(1)\times U(1)]\cr &&
\times U(N)_{eR}.\ee
The subscripts indicate on which generation
multiplet the
$U(N)$s act, $qL$ for the $N=3$ left-handed quarks, $\ell
L$ for the left-handed leptons,
$uR$ for the right-handed up-quarks and
so forth. The kernel of the projection down to the
automorphism group Aut($\aa)$ is
\bb{\rm ker}\,p=
[U(1)\times U(1)&&\!\!\!\!\!\!\!\times U(N)_{qL}
\times U(N)_{\ell L}\times U(N)_{uR}
\times U(N)_{dR}]/[U(1)\times U(1)]\cr &&
\times U(N)_{eR},\ee
and its restrictions to the images of the lifts are
\bb {\rm ker}\,p \cap L({\rm In}(\aa))=\zz_2\times\zz_3,\qq
{\rm ker}\,p \cap \llll(U^n(\aa))=\zz_2\times U(1).\ee
 As a side remark we anticipate that the
maximally extended standard model in noncommutative
geometry \cite{lift}, that gauges 
$U(N)_{uR}$ and $U(N)_{dR}$ simultaneously, is not viable: it
has massless, physical Higgs scalars and also $m_t=m_b$
\cite{gis}. 

The kernel of $i$ is $\zz_2\times U(1)$ in sharp contrast to
the kernel of $\hat L$ which is trivial.
The isospin $SU(2)_L$ and the
colour $SU(3)_c$ are the image of the lift $\hat L$. 
If $q\not=0,$ the image of 
$\ell$ consists of one 
$U(1)\owns w_c=\exp [i\theta ]$ contained in the five
flavour
$U(N)$s. Its embedding depends on $q$: 
\bb \llll(1_2,1,
w_c1_3)=\ell(w_c)=&&\cr 
{\rm diag}\,(&u_11_2\ot1_N\ot1_3,
u_21_2\ot1_N,u_31_N\ot1_3,u_41_N\ot1_3,u_51_N;&\cr 
&\bar u_11_2\ot1_N\ot1_3,
\bar u_21_2\ot1_N,\bar
u_31_N\ot1_3,\bar u_41_N\ot1_3,\bar u_51_N&)\ee
with $u_j=\exp[iy_j\theta ]$ and
\bb 
y_1 = q_{2},\qquad
y_2= -q_{1},\qquad 
y_3 = q_{1}+q_{2} ,\qquad 
y_4 = -q_{1}+q_{2}, \qquad
y_5 = -2q_{1}. \label{y}
\ee
Independently of the embedding, we have indeed {\it
derived} the three conditions (\ref{3cond}), that in the
Yang-Mills-Higgs version had to be imposed. In other words,
in noncommutative geometry the massless electroweak gauge
boson necessarily couples vectorially.

Our goal is now to find the minimal extension $\ell$, that
renders the extended lift symmetric,
$\llll(-u_0,-w)=\llll(u_0,w)$, and that renders
 $\llll(1_2,w)$ single
valued. The first requirement means \{
$ \tilde q_{1}=1$ and
$\tilde q_{2}=0$ \} modulo 2,
with 
\bb \pp{\tilde q_1\cr \tilde q_2}
=\,{\textstyle\frac{1}{3}} \left( \pp{ q_1\cr  q_2}
-\pp{0\cr 1}\right).\ee
 The second requirement means that
$\tilde q$ has integer coefficients.

 The first extension, that comes to mind, has
$q=0$, $\tilde q=\pp{0\cr -1/3}$. With respect to the
interpretation (\ref{innfake}) of the inner
automorphisms, one might object that this is not an
extension at all. With respect to the {\it generic}
characterization (\ref{inntrue}) it certainly is a
non-trivial extension. Anyhow it  fails both tests. The most
general extension, that passes both tests has the form
\bb \tilde q=\pp{2z_1+1\cr 2z_2},\qq
q=\pp{6z_1+3\cr 6z_2+1},\qq
z_1,z_2\in\zz.\ee
Consequently $y_2=-q_1$  cannot vanish, the neutrino comes
out electrically neutral in compliance with the Lorentz force.
Let us normalize the hypercharges to $y_2=-1/2$
and compute the last remaining hypercharge $y_1$,
 \bb
y_1=\,\frac{q_{2}}{2q_{1}}\, 
={\frac{{\textstyle\frac{1}{6}}+z_2 }{
1+2z_1  }}.\ee
We can change the sign of $y_1$ by 
permuting 
$u$ with $d^c$ and $d$ with $u^c$. Therefore it is
sufficient to take $z_1=0,1,2,...$
The minimal such extension, $z_1=z_2=0$, 
recovers nature's choice $y_1={\textstyle\frac{1}{6}}$. 
Its lift, 
\bb \llll(u_0,w)=\rho (u_0,\det w,w)J
\rho (u_0,\det w,w)J^{-1},\ee 
is the fermionic
representation of the standard model considered as
$SU(2)\times U(3)$ Yang-Mills-Higgs model. This lift is
double-valued as the gravitational spin lift (\ref{spi}). The
double-valuedness of
$\llll$ comes from the {\bf discrete} group $\zz_2$ of central
unitaries
$(\pm 1_2,1_3)\ \in\  \zz_2\ \subset\ \Gamma\ \subset\
U^{nc}(\aa) $ and cannot be removed by any central extension
of the form (\ref{ell}).  On the other hand O'Raifeartaigh's
$\zz_2$,
$\pm (1_2, 1_3)\ \in\ \zz_2\ \subset\ U^{nc}(\aa)$ is not a
subgroup of
$\Gamma $. It reflects the symmetry of
$\llll$.

\section{Conclusion}

Central extensions of the lift of automorphisms play three
roles: in the commutative case they unify gravity and
electromagnetism. There, the central unitaries were harmless
and the extensions optional. In general, an extension is
mandatory to reduce the multi-valuedness of the lift and to
reestablish its symmetry which may be lost when suppressing
the harmless unitaries. In the case of the internal space of the
standard model, the non-extended lift is at least 6-valued. The
minimal extension leading to maximal reduction is the key to
the hypercharges of all fermions. The minimal remaining
multi-valuedness is double, precisely as for the lift in
spacetime. This raises the question if the two
double-valuednesses are related and if the internal one
is also accessible to experiment.

Still we wonder: Why does the algebra of the standard model
need a commutative part $\cc$ and produce unitaries
which are harmless and only make a virtual appearance? 
The $\cc$ plays two essential roles. It prints the seat tickets,
that indicate where in the receptacle  Aut$_\hh(\aa)$ of the
lifted automorphisms the harmful unitaries are to be
seated. Its second role is more physical \cite{gis}: Without at
least one (non-vectorial)
$\cc$ in the internal algebra, the symmetry breaking induced
by the spectral action gives identical masses to Dirac
spinors in the same irreducible multiplet.

\vskip 1truecm\noindent
As always, it is a pleasure to acknowledge Bruno Iochum's
constructive critique.


\begin{thebibliography}{47}

\bibitem{book}
 A. Connes, {\it Noncommutative Geometry}, Academic
Press (1994)
\bibitem{grav} A. Connes, {\it Gravity coupled with
matter and the foundation of noncommutative
geometry}, hep-th/9603053, Comm. Math. Phys. 155
(1996) 109
\bibitem{tresch} A. Connes, {\it Noncommutative
geometry and reality},  J. Math. Phys. 36 (1995) 6194
\bibitem{cc} A. Chamseddine \& A. Connes, {\it The
spectral action principle}, hep-th/9606001
\bibitem{florian} F. Girelli, {\it Left-right symmetric models
in noncommutative geometry? } hep-th/0011123
\bibitem{gis} F. Girelli, B. Iochum \& T. Sch\"ucker, {\it
Towards the standard model from noncommutative geometry}
\bibitem{or} L. O'Raifeartaigh, {\it Group Structure of
Gauge Theories}, Cambridge University Press 1986
\bibitem{lift} T. Sch\"ucker, {\it Spin group and almost
commutative geometry}, hep-th/0007047
\bibitem{neu}
H. Rauch, A. Zeilinger, G. Badurek, A. Wilfing, W. Bauspiess \&
U. Bonse, {\it Verification of coherent spinor rotations of
fermions}, Phys. Lett. 54A (1975) 425
\bibitem{ls} S. Lazzarini \& T. Sch\"ucker, {\it Standard
model and unimodularity condition}, hep-th/9801143,
in the proceedings of the Workshop on Quantum
Groups, Palermo, 1997, ed.: Daniel Kastler, Nova
Science, 1999
\bibitem{data}
The Particle Data Group, {\it Particle Physics Booklet}
and ${\tt http://pdg.lbl.gov}$
\bibitem{reb}
R. Asquith, {\it Non-commutative geometry and the 
strong force}, hep-th/9509163, Phys. Lett. B 366 (1996)
220


\end{thebibliography}
 \end{document}